# Destruction of $^3$He in Low Mass Stars and Implications for Chemical Evolution

Corinne Charbonnel

*Laboratoire d'Astrophysique de Toulouse, CNRS, Toulouse, France*

**Abstract.** Recent observations of $^3$He/H in different environments tend to prove that $^3$He is not produced in low mass stars, contrary to the standard predictions of stellar evolution theory. We show how a simple consistent mechanism, namely rotation-induced mixing, can lead to the destruction of $^3$He in low mass stars and simultaneously account for the low $^{12}$C/$^{13}$C ratios and low lithium abundances observed in giant stars of different populations. This process should both naturally account for the recent measurements of $^3$He/H in galactic HII regions and allow for high values of $^3$He observed in some planetary nebulae.

## 1. Failures of the conventional $^3$He evolutionary scenario

Until recently, the main features of $^3$He chemical evolution seemed to be simple, and were mainly dominated by the net production of this light element in low mass stars (stars with masses lower than 2 M$_\odot$). In these objects indeed, initial D is processed to $^3$He during the fully convective pre-main sequence phase. Moreover, while the star is on the main sequence, and due to nuclear reactions of the pp-chains, an $^3$He peak builds up. This peak is engulfed in the convective envelope of the star during the first dredge up on the lower red giant branch. This leads to an increase of the surface abundance of $^3$He. In classical stellar models, $^3$He then survives during the following phases of evolution, and is injected in the ISM by stellar winds and planetary nebulae ejection. In this standard view, $^3$He/H is thus predicted to increase in time in the regions where stellar processing occurs (Rood et al. 1976).

However, galactic chemical evolutionary models including $^3$He production by low mass stars (Vangioni-Flam & Cassé 1995; Olive et al. 1995; Galli et al. 1995; see also Tosi, this volume) are in conflict with recent measurements of $^3$He/H in different environments : They all predict an overproduction of $^3$He (by factors between 5 to more than 20) compared to observations. Actually, the protosolar value $^3$He/H $\simeq (1.5 \pm 0.5) \times 10^{-5}$ (Geiss 1993, Copi et al. 1995) is difficult to reconcile with the values $^3$He/H $= (1-5 \times 10^{-5})$ measured in galactic HII regions (Rood et al. 1995) (low mass stars essentially start to contribute to the ISM enrichment in the era between the birth of the solar system and the present days). These observations essentially leave no room for important production of $^3$He in the Galaxy. This indicates that actually low mass stars do not produce this element.

The situation is even worse when one considers the possible recent detection of D in quasar absorbtion line systems, and if one trusts that the upper limit of $(1.9 \pm 0.5)10^{-4}$ derived from these analysis reflects the primordial D abundance (Songaila et al. 1994; Carswell et al. 1994; Rugers & Hogan 1995; Wampler et al. 1995). Such a high value requires that stellar processing destroys both D and $^3$He in a very efficient way.

The conventional scenario for chemical evolution of $^3$He appears thus to be upset, and the actual contribution of low mass stars to the evolution of $^3$He has to be revised.

## 2. Connection with other chemical anomalies

The problem of $^3$He in low mass stars has to be related with other chemical anomalies which are not accounted for in the framework of the standard stellar theory. Let us focus in particular on $^{12}C/^{13}C$ ratios and $^7$Li abundances which decrease at the stellar surface during the first dredge up phase while the $^3$He abundance increases.

According to the standard scenario, the surface abundances then stay unaltered as the convective envelope slowly withdraws during the end of the red giant branch ascent. However, observations of evolved stars of different populations indicate that the $^{12}C/^{13}C$ ratio and the $^7$Li abundance continue to decrease after the completion of the theoretical first dredge-up (Sneden et al. 1986; Gilroy 1989; Smith & Suntzeff 1989; Brown & Wallerstein 1989; Bellet al. 1990; Gilroy & Brown 1991; Pilachowski et al. 1993; Charbonnel 1994); see Figures 1 and 2.

These data reveal that a non-standard mixing mechanism is acting in low mass stars as they are ascending the red giant branch. More precisely, some major observational constraints provide the evidence that this non-standard mixing occurs above the luminosity at which the hydrogen burning shell crosses the chemical discontinuity created by the convective envelope during the dredge-up (Charbonnel 1994). We now propose a realistic physical process which simultaneously accounts for the observed behavior of $^{12}C/^{13}C$ and $^7$Li in evolved low mass stars, and leads to low $^3$He.

## 3. Rotation-induced mixing on the red giant branch

Among the processes which were proposed to explain the abundance anomalies in evolved stars, rotation-induced mixing is the most promising. We investigated the influence of such a process on the red giant branch (Charbonnel 1995). We used the most recent theoretical developments in the description of the transport of chemicals and angular momentum in stellar interiors, i.e. Zahn's (1992) consistent theory which describes the interaction between meridional circulation and turbulence induced by rotation. In this framework, the global effect of advection moderated by horizontal turbulence can be treated as a diffusion process. In our context, three important points must be emphasized : *1.* The resulting mixing of chemicals in stellar radiative regions is mainly determined by the loss of angular momentum via a stellar wind. *2.* Even in the absence of such mass loss, some mixing can take place wherever the rotation profile presents steep vertical gradients. *3.* Additional mixing is expected near nuclear burning shells. All these conditions are expected to be fulfilled during the non-homologous evolution on the red giant branch.

We estimated how this rotation-induced mixing can modify the chemical abundances. Stellar evolutionary models were computed with the Toulouse code, a version of the Geneva stellar evolution code in which we have introduced the numerical method described in Charbonnel et al. (1992) to solve the diffusion equation. We restricted our study to the case where the stars undergo a moderate wind (see Zahn 1992).

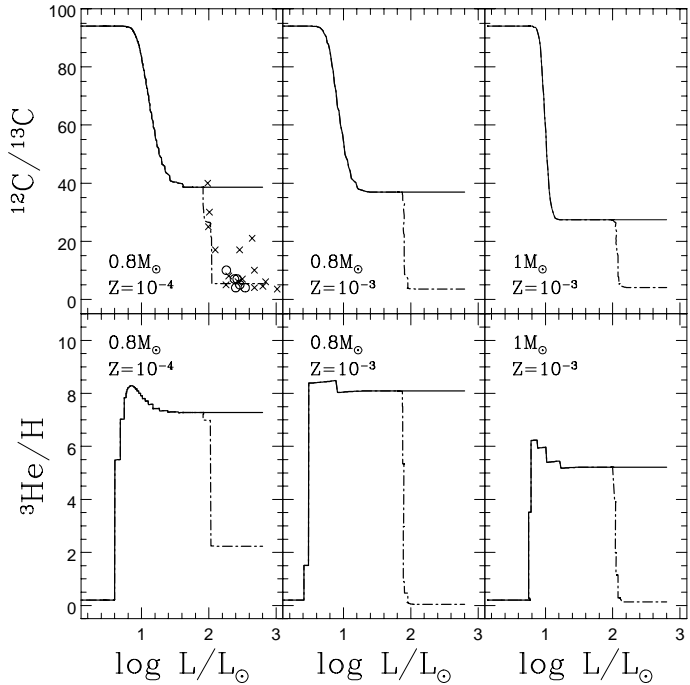

Figure 1. Theoretical behavior of $^{12}C/^{13}C$ and of $^{3}He/H$ in units of $10^{-4}$ as a function of luminosity, for standard evolution (solid lines) and for the evolution including extra-mixing (dashed-dotted lines). Observations of the carbon isotopic ratio in field Population II (crosses; Sneden et al. 1986) and globular cluster M4 (circles; Smith & Suntzeff 1989) giant stars.

Figure 1 shows the influence of rotation-induced mixing on the surface values of $^{12}C/^{13}C$ and $^{3}He$. In the standard case, the post-dilution values of the carbon isotopic ratio and of $^{3}He/H$ remain constant. The $^{12}C/^{13}C$ is then substantially higher than observed.

However, when extra-mixing begins to act, the $^{12}C/^{13}C$ and $^{3}He/H$ both rapidly drop. The observed behavior of $^{12}C/^{13}C$ is well reproduced, and one reaches the low values currently observed in globular cluster giants, namely 3-8.

Simultaneously, when $^{3}He/H$ diffuses, it rapidly reaches the region where it is nuclearly burned by the $^{3}He(\alpha,\gamma)^{7}Be$ reaction. This leads to a rapid decrease of the surface value of $^{3}He/H$, confirming the predictions by Hogan (1995). As can be seen in Figure 2, lithium is also rapidly transported from the convective envelope down to the region where it is burned by proton capture. Due to rotation-induced mixing on the RGB, the surface abundance of lithium rapidly decreases down to the very low values observed in the halo giants.

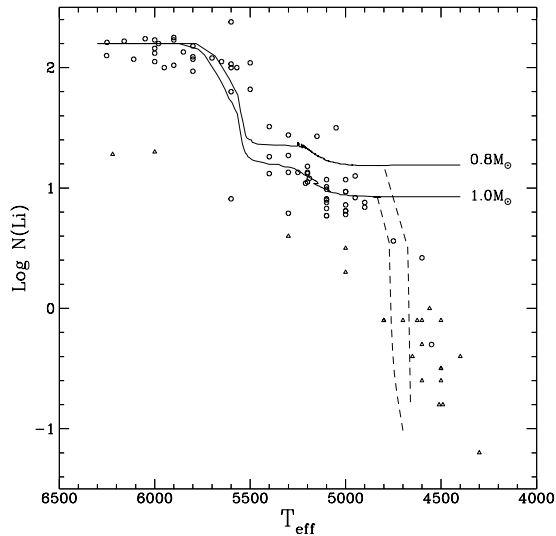

Figure 2. Theoretical behavior of the lithium abundance as a function of $T_{eff}$, for standard evolution (solid lines) and for the evolution including extra-mixing (dashed-dotted lines), for 0.8 and $1M_\odot$ models computed with $Z=10^{-4}$. The very low lithium abundances observed in the most evolved stars of the sample (halo giant stars from Pilachowski et al. (1993); open circles for real lithium detection, open triangles for upper limits) can only be reproduced when rotation-induced mixing is taken into account.

## 4. Conclusions

We showed how a simple consistent mechanism, based on the most recent developments in the description of rotation-induced mixing, simultaneously accounts for the low $^{12}C/^{13}C$ ratios and $^7Li$ abundances in red giant stars and naturally leads to the destruction of $^3He$ in low mass stars.

Let us stress an important point concerning the high value of $^3He/H$ observed in a few Planetary Nebulae. Observations of $^{12}C/^{13}C$ ratios in M67 evolved stars (Gilroy & Brown 1991) show that the extra-mixing process described above is only efficient when the hydrogen burning shell has crossed the discontinuity in molecular weight built by the convective envelope during the first dredge-up (Charbonnel 1994). Before this evolutionary point, the mean molecular weight gradient probably acts as a barrier to the mixing in the radiative zone. Above this point, no gradient of molecular weight exists anymore above the hydrogen burning shell, and extra-mixing is free to act. It is crucial to note that for stars which do not undergo the helium flash (i.e. stars with masses higher than 1.7 to 2 $M_\odot$), the hydrogen burning shell does not have time to reach the chemically homogeneous region during the short ascent of the RGB, and extra-mixing does not occur (This point is confirmed by observations of $^{12}C/^{13}C$ in giant stars of galactic clusters whose turn off masses are higher than 2 $M_\odot$). So if the high value of $^3He/H$ observed in the PN NGC 3242 (Rood et al. 1992) is confirmed, it requires in our framework that the progenitor of this object had an initial mass higher than 1.7 - 2 $M_\odot$.

Our model leads to a strong revision of the actual contribution of low mass stars to $^3He$ chemical evolution. It can simultaneously account for the recent measurements of $^3He/H$ in different environments and allow for the high value of $^3He$ observed in some planetary nebulae.